\documentclass[12pt]{article}
\makeatletter
\newcommand{\singlespacing}{\let\CS=\@currsize\renewcommand{\baselinestretch}{1}\tiny\CS}
\usepackage{epsfig}
\usepackage{amsbsy}
\usepackage{amssymb,color}
\usepackage[table]{xcolor}
\begin{document}
\baselineskip=24pt
%\singlespacing
%\doublespacing
\parskip = 10pt
\def \qed {\hfill \vrule height7pt width 5pt depth 0pt}
\newcommand{\ve}[1]{\mbox{\boldmath$#1$}}
\newcommand{\IR}{\mbox{$I\!\!R$}}
\newcommand{\1}{\Rightarrow}
\newcommand{\bs}{\baselineskip}
\newcommand{\esp}{\end{sloppypar}}
\newcommand{\be}{\begin{equation}}
\newcommand{\ee}{\end{equation}}
\newcommand{\beanno}{\begin{eqnarray*}}
\newcommand{\inp}[2]{\left( {#1} ,\,{#2} \right)}
\newcommand{\eeanno}{\end{eqnarray*}}
\newcommand{\bea}{\begin{eqnarray}}
\newcommand{\eea}{\end{eqnarray}}
\newcommand{\ba}{\begin{array}}
\newcommand{\ea}{\end{array}}
\newcommand{\nno}{\nonumber}
\newcommand{\dou}{\partial}
\newcommand{\bc}{\begin{center}}
\newcommand{\ec}{\end{center}}
\newcommand{\2}{\subseteq}
\newcommand{\cl}{\centerline}
\newcommand{\ds}{\displaystyle}
\def\refhg{\hangindent=20pt\hangafter=1}
\def\refmark{\par\vskip 2.50mm\noindent\refhg}

\title{\sc Univariate and Bivariate Geometric Discrete Generalized Exponential Distributions}

\author{Debasis Kundu$^{1}$ \& Vahid Nekoukhou$^{2}$}

\date{}
\maketitle

\begin{abstract}
Marshall and Olkin (1997, Biometrika, 84, 641 - 652) introduced a very powerful method to introduce an additional parameter
to a class of continuous distribution functions and hence it brings more flexibility to the model.  They have demonstrated their
method for the exponential and Weibull classes.  In the same paper they have briefly indicated regarding its bivariate extension.
The main aim of this paper is to introduce the same method, for the first time, to the class of discrete generalized exponential distributions
both for the univariate and bivariate cases.  We investigate several properties of the proposed univariate and bivariate classes.
The univariate class has three parameters, whereas the bivariate class has five parameters.  It is observed that depending on the
parameter values the univariate class can be both zero inflated as well as heavy tailed.  
We propose to use EM algorithm to estimate
the unknown parameters.  
Small simulation experiments have been 
performed to see the effectiveness of the proposed EM algorithm, and a bivariate data set has been analyzed and it is observed that 
the proposed models and the EM algorithm
work quite well in practice.

\end{abstract}



\noindent$1$Department of Mathematics and Statistics, Indian Institute of Technology Kanpur, Kanpur,
Pin 208016, India.  \ \ e-mail: kundu@iitk.ac.in.

\noindent$^2$ Department of Statistics, Khansar Faculty of Mathematics and Computer Science, Khansar, Iran.

\newpage

\section{\sc Introduction}

Generalized exponential (GE) distribution originally introduced by Gupta and Kundu (1999) has received considerable
attention in recent years.  It is an absolute continuous univariate distribution with several interesting properties.  It
has been used quite successfully as an alternative to a gamma or a Weibull distribution.  Although, often in practice we
use a continuous random variable mainly due to analytical tractability, discrete data occur in practice
quite naturally in various fields.  For example, the number of deaths due to a particular cause in a place during a month,
the number of attempts needed to crack a computer password or the number of goals scored by a particular team are
purely discrete in nature.  In these cases it is better to
analyze these data using a discrete probability model rather than a continuous probability model.  Several attempts have been
made to generate various discrete probability distributions and to develop their properties, see for example the book by
Johnson et al. (2005), and the references cited therein.

Recently, Nekoukhou et al. (2013) introduced a discrete generalized exponential (DGE) distribution, which can be considered as
the discrete analogue of the absolute continuous GE distribution of Gupta and Kundu (1999).  The DGE distribution proposed by
Nekoukhou et al.
(2013) is a very flexible two-parameter distribution.  The probability mass function of the DGE distribution can be a decreasing
or a unimodal function.  Similarly, hazard function of the DGE distribution can be increasing, decreasing or constant depending
on the shape parameter.  Hence, the geometric distribution can be obtained as a special case of the DGE distribution.  A DGE model is appropriate for analyzing both over and under-dispersed discrete data, in view of the fact that the variance can be larger or smaller than the mean. It has been used
to analyze various discrete data sets, and the performances were quite satisfactory.

Bivariate distributions are mainly used to analyze the marginals and also to model the dependence structure between the two
marginals.  Extensive work has been done to propose different bivariate continuous distributions and to develop their properties.
Similar to the continuous univariate data, the continuous bivariate data are also analyzed quite often in practice mainly due to analytical tractability. See for example the book by Balakrishnan and Lai (2009) and the references cited therein for
different continuous bivariate probability distributions and for their various properties and applications.

Discrete bivariate data also occur quite naturally in practice.  For example, the number of goals scored by two competing teams
or the number of insurance claims for two different causes are purely discrete in nature.  Several bivariate discrete distributions
have been proposed in the statistical literature mainly to analyze bivariate discrete data.  Recently, Lee and Cha (2015) introduced
two general classes of discrete bivariate distributions, and Nekoukhou and Kundu (2017) proposed a four-parameter
 bivariate discrete generalized exponential distribution.  See also the books by
Kocherlakota and Kocherlakota (1992), Johnson et al. (1997) and the numerous references cited there in this respect.

Marshall and Olkin (1997) introduced a very efficient mechanism to introduce an extra parameter to a class of continuous univariate
distribution functions and hence it brings more flexibility to the existing model.  They have illustrated their methods by using
exponential and Weibull distribution functions.  Since then an extensive amount of work has been done for generalizing several univariate
classes of distribution functions, see for example Adamidis and Loukas (1998), Louzada et al. (2014), Ristic and Kundu (2015,
2016) and the references cited therein.  Marshall and Olkin (1997) also mentioned about the extension of their method to the
bivariate case.  Recently, Kundu and Gupta (2014) and Kundu (2015) applied that method for the bivariate Weibull and bivariate
generalized exponential distributions, respectively.

Although an extensive amount of work has been done for continuous distributions particularly for univariate case, no attempt has been
made for the discrete distribution except the work by G\'{o}mez-D\'{e}niz (2010).  G\'{o}mez-D\'{e}niz (2010) adopted the same method as of
Marshall and Olkin (1997) and proposed a new generalized version of the geometric distribution.  No work has been done, at least
not known to the authors about the bivariate case.

The aim of this paper is two fold.  First we introduce the univariate geometric discrete generalized exponential (GDGE)
distribution.  We develop several properties of the proposed univariate GDGE distribution.  The proposed univariate GDGE
distribution has three parameters.  The probability mass function (PMF) of a univariate GDGE distribution can take variety of
shapes.  It can be zero inflated as well as heavy tailed.  It may be mentioned that not too many univariate discrete distributions
have these properties.  Then we introduce the bivariate GDGE distribution.  It  has five parameters.  Due to the presence of five
parameters the bivariate GDGE distribution is a very flexible bivariate discrete distribution. Its marginals are univariate
GDGE distributions.  We develop several properties of the bivariate GDGE distribution.  We provide various dependency measures
also.  The maximum likelihood estimators (MLEs) of the unknown parameters cannot be obtained in closed forms.  One needs to solve
five non-linear equations to compute the MLEs of the unknown parameters.  We have proposed to use EM algorithm to compute the
MLEs of the unknown parameters.  
Small simulation experiments have been 
performed to see the effectiveness of the proposed EM algorithm.
We have analyzed two univariate and one bivariate data sets to illustrate how the method can be used in practice.  
It is observed that the performances of the models and the proposed EM algorithm work quite satisfactory.

%\enlargethispage{1.00 cm}

The rest of the paper is organized as follows.  In Section 2 we provide a brief background of the GE and DGE
distributions.  We introduce and discuss several properties of the univariate GDGE distribution in Section 3.  In Section 4 we
discuss about the bivariate GDGE distribution.  In Section 5, we consider different inferential issues for both the univariate
and bivariate cases.  Simulation results and the data analyses have been presented in Section 6.  Finally we conclude the paper in Section 7.

\section{\sc Preliminaries}

\subsection{\sc The GE Distribution}

The absolute continuous GE distribution was proposed by Gupta and Kundu (1999) as an alternative to the well known gamma and Weibull distributions.  The
two-parameter GE distribution has the following probability density function (PDF), cumulative distribution function (CDF), and the
hazard rate function,  respectively;
\bea
f_{GE}(x; \alpha, \lambda) & = & \alpha \lambda e^{-\lambda x} (1 - e^{-\lambda x})^{\alpha-1}; \ \ \ x > 0,  \label{pdf-ge}  \\
F_{GE}(x; \alpha, \lambda) & = & (1 - e^{-\lambda x})^{\alpha}; \ \ \ x > 0,   \label{cdf-ge}   \\
h_{GE}(x; \alpha, \lambda) & = & \frac{\alpha \lambda e^{-\lambda x} (1 - e^{-\lambda x})^{\alpha-1}}{1 - (1 - e^{-\lambda x})^{\alpha}};
\ \ \ \ x > 0.  \label{hf-ge}
\eea
Here $\alpha > 0$ and $\lambda > 0$ are the shape and the scale parameters, respectively.  From now on a GE distribution with the shape parameter $\alpha$
and the scale parameter $\lambda$ will be denoted by GE$(\alpha,\lambda)$.  Similar to the gamma and Weibull distributions,
it is also a generalization of the exponential distribution and hence, exponential distribution can be obtained as a special case.
The PDF (\ref{pdf-ge}) and the hazard rate function (\ref{hf-ge}) of a GE distribution can take various shapes.  The PDF can be a
decreasing or a unimodal function and the hazard rate function can be an increasing, a decreasing or a constant function  depending
on the shape parameter.  It has been observed by several authors during the last fifteen years that this model can be used quite
effectively as an alternative to the Weibull and gamma distributions for many practical problems.  Interested readers are referred to
the review articles by Gupta and Kundu (2007) and Nadarajah (2011) and the recently published monograph by Al-Hussaini and Ahsanullah
(2015).

\subsection{\sc The DGE Distribution}

Recently, the DGE distribution was proposed by Nekoukhou et al. (2013).  A discrete random variable $X$ is said to have a DGE
distribution with parameters $\alpha$ and $\ds p$ $\ds (= e^{-\lambda})$, if the PMF of $X$ can be written as follows:
\be
f_{DGE}(x; \alpha,p) = P(X = x) = (1 - p^{x+1})^{\alpha} - (1-p^x)^{\alpha}, \ \ \  \ \ \ x \in \mathbb{N}_0 = \{0, 1, 2, \ldots \}.
\label{dge-pmf}
\ee
The corresponding CDF becomes
\be
F_{DGE}(x; \alpha, p) = P(X \le x) = \left \{ \matrix{0 & \hbox{if} & x < 0  \cr (1 - p^{[x]+1})^{\alpha} & \hbox{if} & x \ge 0.} \right .
   \label{dge-cdf}
\ee
Here $[x]$ denotes the largest integer less than or equal to $x$.  From now on a DGE distribution with parameters $\alpha$ and $p$
will be denoted by DGE$(\alpha,p)$.  The PMF and the hazard rate function of a DGE distribution can take various shapes.
The PMF can be a decreasing or a unimodal, and the hazard rate function can be an increasing or a decreasing function.
Different moments and the distribution of the order statistics were obtained
by the authors in the same paper. A DGE model is appropriate for modeling both over and under-dispersed data since, in this model, the variance can be
larger or smaller than the mean which is not the case with most of the standard classical discrete distributions.

The following representation of a DGE random variable becomes very useful.  Suppose $X \sim$ DGE$(\alpha,p)$, then for
$\ds \lambda = -\ln p$,
\be
Y \sim \hbox{GE}(\alpha,\lambda) \Longrightarrow X = [Y] \sim \hbox{DGE}(\alpha, p).    \label{repre}
\ee
Using (\ref{repre}), the generation of a random sample from a DGE$(\alpha,p)$ becomes very simple.  For example, first we can generate
a random sample $Y$ from a GE$(\alpha,\lambda)$ distribution, and then by considering $X = [Y]$, we can obtain a generated sample from
 DGE$(\alpha,p)$.

\section{\sc Univariate GDGE Distribution}

%\subsection{\sc Definition and Interpretations}

Suppose $X_1, X_2, \ldots$ are independent identically distributed (i.i.d.) DGE$(\alpha, p)$ random variables, where $0 < \alpha
< \infty$ and $0 < p < 1$ and $N$ is a geometric random variable with the following PMF for $0 < \theta < 1$,
\be
P(N=n) = \theta (1-\theta)^{n-1}; \ \ \ \ n\in \mathbb{N}=\{1,2,...\}.    \label{pmf-geom}
\ee
The above geometric distribution will be denoted by GM$(\theta)$, in the rest of the paper.  It is further assumed that $N$ is
independent of $X_i$'s.  Let us define a new random variable
\be
X = \max\{X_1, \ldots, X_N\}.
\ee
Then the distribution of $X$ is said to have the univariate GDGE distribution and it will be denoted by UGDGE$(\alpha, p, \theta)$.

Note that the univariate GDGE distribution can be used quite effectively to analyze a parallel system with random number of
components.  In this case the number of components is a random quantity and it follows a geometric distribution where as the
lifetime of each component follows a DGE distribution.  It may be mentioned that the analysis of a parallel system with random
number of components has received considerable attention for quite sometime in the statistical reliability literature,
see for example Bartoszewicz (2001), Shaked and Wong (1997) and the references cited therein.
%\enlargethispage{1 cm}

\begin{figure}[h]
\begin{center}
\includegraphics[height=6cm,width=6cm]{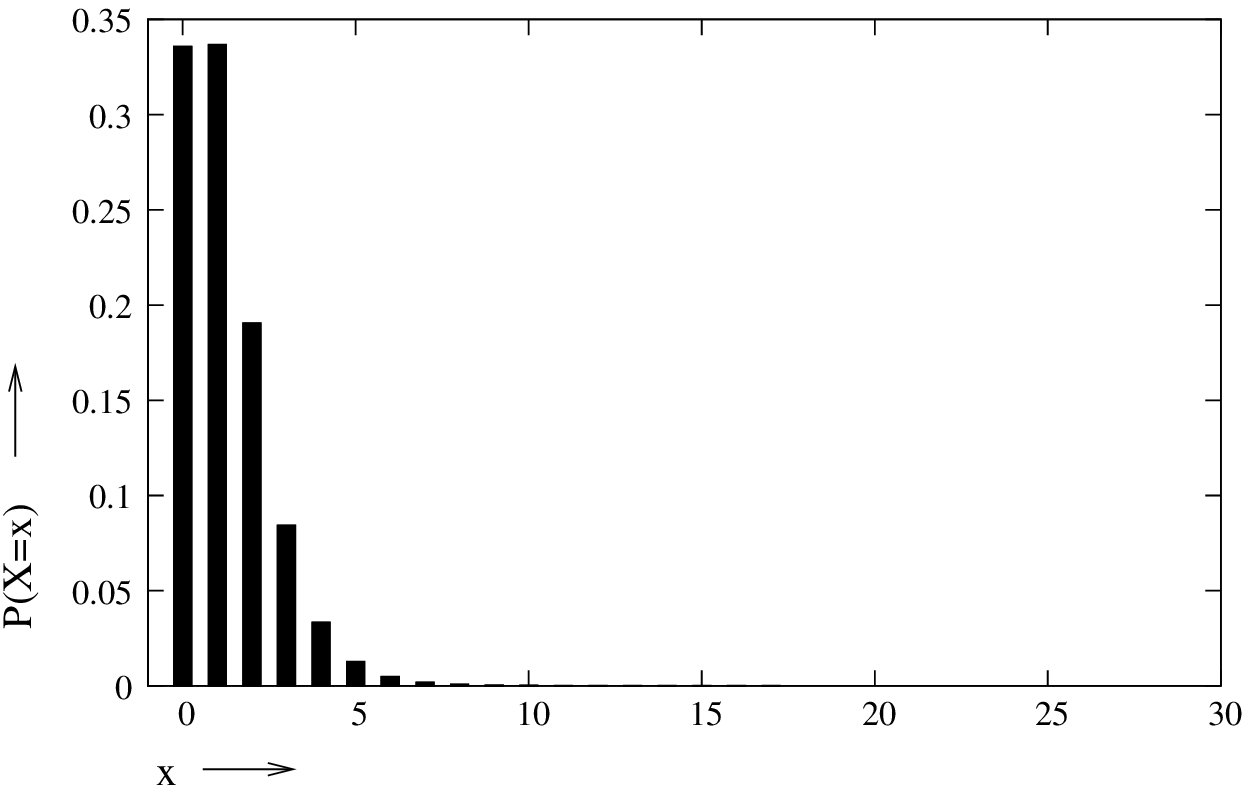}
\caption{The PMF of a univariate GDGE distribution when $\alpha$ = 1.5, $\theta$ = 0.5, $p = e^{-1.0}$.  \label{pmf-1}}
\includegraphics[height=6cm,width=6cm]{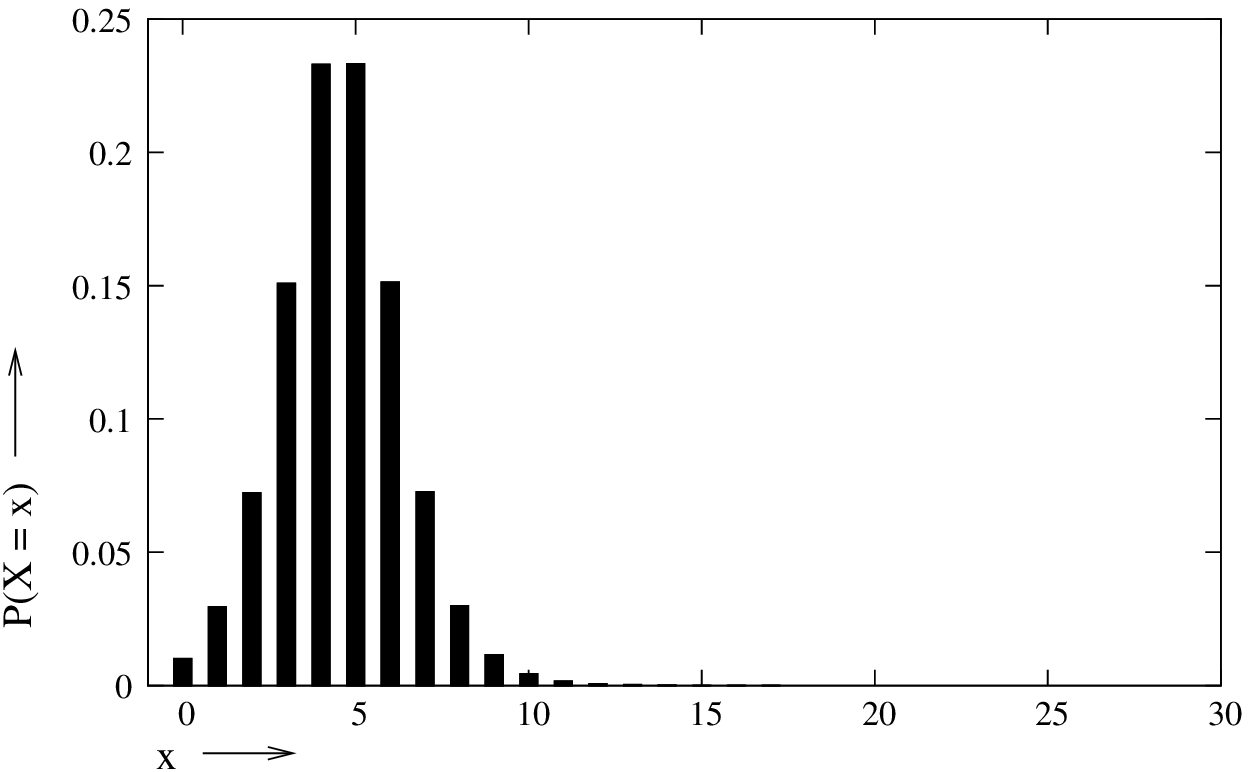}
\caption{The PMF of a univariate GDGE distribution when $\alpha$ = 1.5, $\theta$ = 0.01, $p = e^{-1.0}$.   \label{pmf-2}}
\includegraphics[height=6cm,width=6cm]{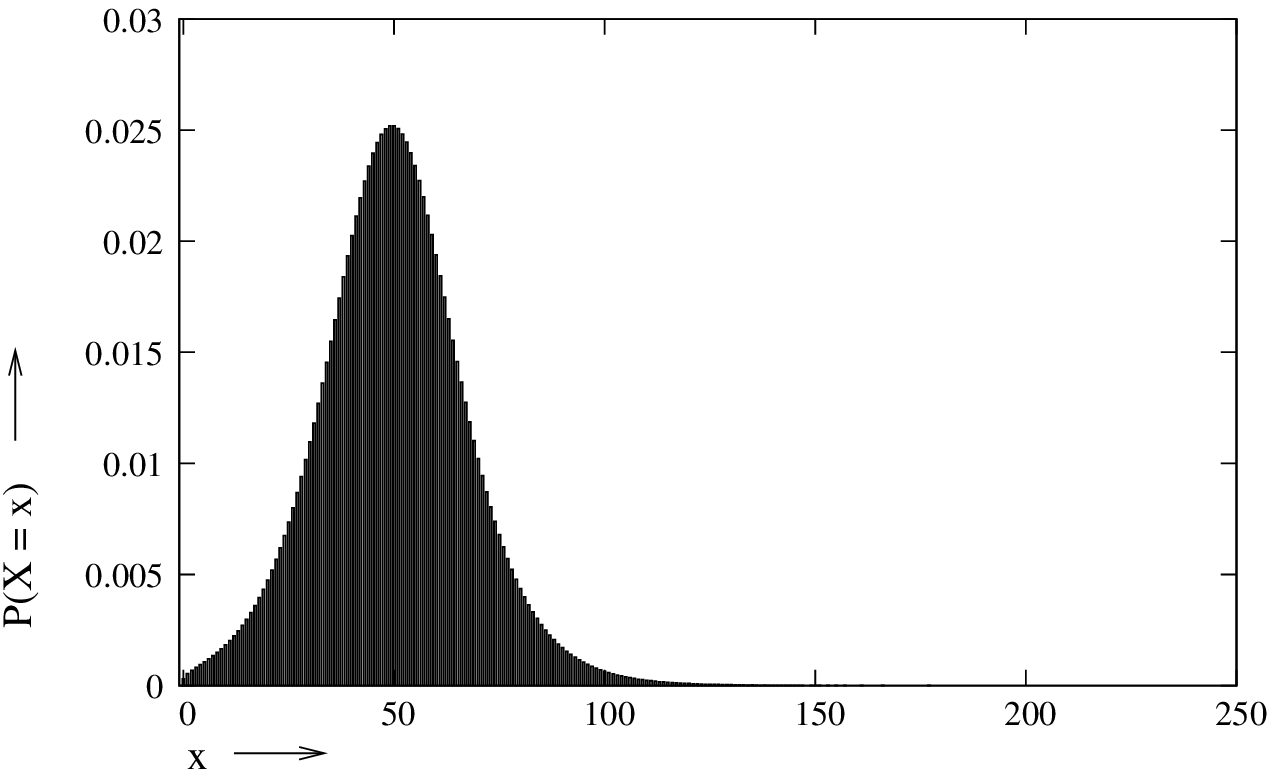}
\caption{The PMF of a univariate GDGE distribution when $\alpha$ = 1.5, $\theta$ = 0.01, $p = e^{-0.1}$.   \label{pmf-3}}
\end{center}
\end{figure}

The CDF of $X$ for $x \ge 0$, can be obtained as follows.
\bea
F(x)  =  P(X \le x) & = & \sum_{n=1}^{\infty} P(X \le x, N = n) = \sum_{n=1}^{\infty} P(X \le x|N = n)P(N = n)    \nonumber  \\
& = & \theta (1 - p^{[x]+1})^{\alpha} \sum_{n=0}^{\infty} (1 - p^{[x]+1})^{n \alpha} (1-\theta)^{\alpha}  \nonumber  \\
& = & \frac{\theta (1 - p^{[x]+1})^{\alpha}}{1 - (1-\theta) (1 - p^{[x]+1})^{\alpha}}.    \label{u-cdf-1}
\eea
Hence, the PMF of $X$ becomes
\be
P(X = x) = f_X(x) =  \frac{\theta \left [ (1 - p^{x+1})^{\alpha} - (1 - p^x)^{\alpha} \right ]}{\left [ 1 - (1-\theta)(1 - p^{x+1})^{\alpha}
\right ]\left [1 - (1-\theta)(1 - p^x)^{\alpha} \right ]}; \ \ \ \ x \in \mathbb{N}_0.    \label{univ-pmf}
\ee
It is clear that if $\theta$ = 1, then univariate GDGE becomes univariate DGE distribution, and
if   $\theta$ = $\alpha$ = 1, then it becomes a geometric distribution.  Therefore,
clearly the proposed univariate GDGE distribution is a generalization of the geometric and univariate DGE distributions.  In fact from
(\ref{u-cdf-1}) it is clear that for fixed $x$ as $\theta \rightarrow 0$, then $F(x) \rightarrow 0$.  Hence, for any fixed $x$,
as $\theta \rightarrow 0$, $P(X > x) \rightarrow 1$.  It implies that
as $\theta \rightarrow 0$, the univariate GDGE becomes a heavy tailed distribution.  In Figures \ref{pmf-1} to \ref{pmf-3} we have
provided the plots of the PMFs of the univariate GDGE distribution for different parameter values.  It is clear from the plots of the PMFs
that it can take different shapes depending on the parameter values.  It can be a decreasing or a unimodal function, and it can be heavy  tailed also.

Note that the PMF of $X$ for $x \in \mathbb{N}_0$ can be written as
\begin{eqnarray}
f_X(x)=F_X(x)-F_X(x-1)=w(x)f_{DGE}(x;\alpha,p),
\end{eqnarray}
where
$$
w(x)=\frac{\theta}{\left[1-(1-\theta)F_{DGE}(x;\alpha,p)\right]\left[1-(1-\theta)F_{DGE}(x-1;\alpha,p)\right]}.
$$
Therefore, the PMF of $X$ can be written as the weighted DGE probability mass functions with the weight function $w(x)$.
The hazard rate function of $X$ is also given by
\begin{eqnarray}
h_X(x)=w^*(x)h_{DGE}(x;\alpha,p),
\end{eqnarray}
where
$$w^*(x)=\frac{\theta}{1-(1-\theta)F_{DGE}(x-1;p,\alpha)},$$
where $h_{DGE}$ denotes the hazard rate function of a DGE distribution.

The $\gamma$-th percentile point of a GDGE$(\alpha,\theta,p)$ distribution is given by
\begin{eqnarray}\nonumber
\xi_\gamma=\frac{\ln\left\{1-[\frac{\gamma}{\theta+\gamma(1-\theta)}]^{1/\alpha}\right\}}{\ln p}-1.
\end{eqnarray}
Now we will show that the univariate GDGE can be written as an infinite mixture of DGE distributions.

\noindent {\sc Theorem 1:} The PMF of a UGDGE$(\alpha,\theta,p)$  distribution can be written as an infinite mixture
of DGE distributions for $0 < \theta < 1$.

\noindent {\sc Proof:} By using the series representation
$$
(1-z)^{-1} = \sum_{k=0}^{\infty} z^k; \ \ \ \ \hbox{for} \ \ \ |z| < 1,
$$
the CDF of a UGDGE$(\alpha,\theta,p)$ can be written as
\be
F(x) = \theta \sum_{k=0}^{\infty} (1-\theta)^k(1-p^{[x]+1})^{\alpha(k+1)}.    \label{u-cdf}
\ee
Therefore,
$$
P(X = x) = \theta \sum_{k=0}^{\infty} (1-\theta)^k \left \{(1-p^{x+1})^{\alpha(k+1)} - (1-p^{x})^{\alpha(k+1)} \right \}; \ \ \
x \in \mathbb{N}_0.
$$
Hence, the result follows.   \qed

\noindent Now using the probability generating function (PGF) and the moment generating function (MGF)
of a DGE$(\alpha,p)$, see Nekoukhou and Kundu (2017), the PGF and MGF of
UGDGE$(\alpha,\theta,p)$ can be easily obtained as
$$
G_X(z) = \theta \sum_{k=0}^{\infty} \sum_{j=1}^{\infty} (-1)^j {\alpha(k+1)\choose{j}} (1-\theta)^k \frac{1-p^j}{1-zp^j}; \ \ \ \ |z| < 1,
$$
and
$$
M_X(t)=\theta\sum_{j=1}^\infty\sum_{n=1}^\infty (-1)^{j+1}{n\alpha \choose j}\frac{(1-p^j)(1-\theta)^{n-1}}{1-p^je^t}, \quad t<-\ln p,
$$
respectively.  Both the proofs can be easily obtained, and hence the details are avoided.
It must be mentioned that for an integer $\alpha$,
$\sum_{j=1}^\infty$ should be replaced by $\sum_{j=1}^{\alpha}$; see Nekoukhou et al. (2013).

Different moments of the univariate GDGE distribution can be obtained as follows.  If $X \sim$ UGDGE$(\alpha,\theta,p)$, then
the $r$th moment of $X$ can be obtained as
\bea
E(X^r) = \sum_{x=1}^{\infty} \{x^r - (x-1)^r\}P(X \ge x) = \sum_{x=1}^{\infty} \{x^r - (x-1)^r\} \frac{1 - (1-p^x)^{\alpha}}
{1-(1-\theta)(1-p^x)^{\alpha}}.
\eea
Therefore, in particular the mean and the second moment of $X$ become
\beanno
E(X) & = & \sum_{x=1}^{\infty} \frac{1 - (1-p^x)^{\alpha}}{1-(1-\theta)(1-p^x)^{\alpha}}
\eeanno
and
\beanno
E(X^2) & = & \sum_{x=1}^{\infty} (2x-1)\frac{1 - (1-p^x)^{\alpha}}{1-(1-\theta)(1-p^x)^{\alpha}}.
\eeanno
We have the following results regarding the stochastic ordering of the family of univariate GDGE distributions.

\noindent {\sc Result 1:} If $X_1 \sim$ UGDGE$(\alpha_1,\theta,p)$ and $X_2 \sim$ UGDGE($\alpha_2, \theta, p)$, then
for $\alpha_1 > \alpha_2$, $X_1$ is stochastically larger than $X_2$.

\noindent {\sc Result 2:} If $X_1 \sim$ UGDGE$(\alpha,\theta,p_1)$ and $X_2 \sim$ UGDGE($\alpha, \theta, p_2)$, then
for $p_1 >  p_2$, $X_1$ is stochastically larger than $X_2$.

\noindent {\sc Result 3:} If $X_1 \sim$ UGDGE$(\alpha,\theta_1,p)$ and $X_2 \sim$ UGDGE($\alpha, \theta_2, p)$, then
for $\theta_1 < \theta_2$, $X_1$ is stochastically larger than $X_2$.

\noindent {\sc Proof:} The proofs of Results 1 \& 2 can be obtained using (\ref{u-cdf}), and the proof of
Result 3 can be obtained using (\ref{u-cdf-1}).    \qed

Let us recall the Marshall-Olkin generalized exponential (MOGE) distribution as introduced by Ristic and Kundu (2015).  Suppose
$Y_1, Y_2, \ldots$ are i.i.d. GE$(\alpha, \lambda)$ random variables, and $N$ is a geometric random variable with the PMF as defined
in (\ref{pmf-geom}).  It is also assumed that $N$ and $Y_i$'s are independently distributed.  Then
$$
Y = \max\{Y_1, \ldots, Y_N\}
$$
is said to have MOGE distribution with the parameters $\alpha$ and $p = e^{-\lambda}$, and it will be denoted by
MOGE$(\alpha,\theta,p)$.  Now we have the following result.

\noindent {\sc Theorem 2:} If $Y \sim$ MOGE$(\alpha, \theta, p)$, then $X = [Y] \sim$ UGDGE$(\alpha, \theta, p)$.

\noindent {\sc Proof:} Since,
$$
P(Y < y) = P(Y \le y) = \theta \sum_{j=1}^{\infty} (1-p^y)^{\alpha j} (1-\theta)^j = \frac{\theta (1-p^y)^{\alpha}}{1-(1-\theta)(1-p^y)^{\alpha}},
$$
therefore, for $y \in \mathbb{N}_0$
$$
P(X \le y) = P(Y < y+1) = \frac{\theta (1-p^{y+1})^{\alpha}}{1-(1-\theta)(1-p^{y+1})^{\alpha}}.
$$
Hence, for $y \ge 0$,
$$
P(X \le y) = \frac{\theta (1-p^{[y]+1})^{\alpha}}{1-(1-\theta)(1-p^{[y]+1})^{\alpha}}.
$$
\qed

\noindent Theorem 2 can be used quite effectively to generate samples from a UGDGE$(\alpha, \theta, p)$.  The following
algorithm can be used for that purpose.

\noindent {\sc Algorithm 1:}
\begin{enumerate}

\item First generate $N$ from a GM$(\theta)$.

\item Generate $Y$ from GE$(N \alpha, -\ln p)$ using the inverse transformation method.

\item Obtain the required random variable $X$ as $X = [Y]$.

\end{enumerate}

Now, we will show that the univariate GDGE distributions are closed under geometric maximum.
More precisely, we have the following result.

\noindent {\sc Theorem 3:} Let $\{U_i: i\geq 1\}$ be a sequence of i.i.d. UGDGE$(\alpha,\theta,p)$ random variables,
and $M\sim$ GM$(q)$, $0<q<1$. In addition, suppose that $U_i$'s and $M$ are independent, then we have the following result
\begin{eqnarray}
U=\max \{U_1,U_2,...,U_M\} \sim {\rm UGDGE} (\alpha, q \theta, p).
\end{eqnarray}
\noindent {\sc Proof:} Note that
\beanno
P(U\leq u)&=&\sum_{m=1}^\infty P(U_1\leq u,...,U_M\leq u|M=m) P(M = m)  \\
&=&q\sum_{m=1}^\infty \left[\frac{\theta(1-p^{u+1})^{\alpha}}{1-(1-\theta)(1-p^{u+1})^{\alpha_1}}\right]^m(1-q)^{m-1}\\
&=&\frac{q \theta (1-p^{u+1})^{\alpha}}{1-(1-q \theta)(1-p^{u+1})^{\alpha}}.
\eeanno     \qed

The following results will be useful for future development, mainly for developing the EM algorithm to compute the
MLEs of the unknown parameters.  First, note that the joint PMF of $(X, N)$, say $f_{X,N}$, is given by
\begin{eqnarray}
f_{X,N}(x,n)=\theta(1-\theta)^{n-1}\left[(1-p^{x+1})^{n\alpha}-(1-p^{x})^{n\alpha}\right], \quad x\in\mathbb{N}_0, n\in\mathbb{N}.
\end{eqnarray}
Therefore, the joint CDF of $(X,N)$ is also given by
\begin{eqnarray}\nonumber
F_{X,N}(x,n)&=&\sum_{j=1}^n P(X\leq x|N=j)P(N=j)\\\nonumber
&=&\theta\sum_{j=1}^n(1-p^{[ x] +1})^{j\alpha}(1-\theta)^{j-1}\\
&=&\frac{\theta(1-p^{[ x] +1})^{\alpha}\left[1-(1-p^{[ x] +1})^{n\alpha}(1-\theta)^n\right]}{1-(1-\theta)(1-p^{[ x] +1})^{\alpha}}.
\end{eqnarray}
Clearly, the CDF of $X$ also can be obtained as follows
\begin{eqnarray}
F_X(x)=\displaystyle \lim_{n\rightarrow \infty} F_{X,N}(x,n)=\frac{\theta(1-p^{[ x] +1})^{\alpha}}{1-(1-\theta)(1-p^{[ x] +1})^{\alpha}}.
\end{eqnarray}
The conditional PMF of $N$ given $X=x$ is given by
\begin{eqnarray}\nonumber
f_{N|X}(n|x)&=&P(N=n|X=x)\\\nonumber
&=&(1-\theta)^{n-1}\frac{f_{DGE}(x;n\alpha,p)}{f_{DGE}(x;\alpha,p)}\\
&\times&\left[1-(1-\theta)F_{DGE}(x;\alpha,p) \right]\left[1-(1-\theta)F_{DGE}(x-1;\alpha,p) \right].
\end{eqnarray}

The conditional expectation of $N$ can be obtained as
\begin{eqnarray}\nonumber
E(N|X=x)&=&\frac{F_{DGE}(x;\alpha,p)\left[1-(1-\theta)F_{DGE}(x-1;\alpha,p) \right]}{f_{DGE}(x;\alpha,p)\left[1-(1-\theta)F_{DGE}(x;\alpha,p) \right]}\\
&-&\frac{F_{DGE}(x-1;\alpha,p)\left[1-(1-\theta)F_{DGE}(x;\alpha,p) \right]}{f_{DGE}(x;\alpha,p)\left[1-(1-\theta)F_{DGE}(x-1;\alpha,p) \right]}.
\end{eqnarray}

\section{\sc Bivariate GDGE Distribution}

Suppose $X_1, X_2, \ldots$ are i.i.d. DGE$(\alpha_1,p_1)$ random variables, $Y_1, Y_2, \ldots $ are i.i.d DGE$(\alpha_2,p_2)$ random
variables and $N$ is a GM$(\theta)$ random variable.  All the random variables are independently distributed.
Now consider the bivariate discrete random variable $(X,Y)$, where
\begin{eqnarray}
X=\max\{X_1,X_2,...,X_N\} \quad {\rm and} \quad Y=\max\{Y_1,Y_2,...,Y_N\}.
\end{eqnarray}
The joint distribution of $(X,Y)$ is said to have the bivariate GDGE distribution. The following interpretations can be provided
for a bivariate GDGE model.

\noindent {\sc Parallel Systems:} Consider two systems, say 1 and 2, each having $N$ number of independent and identical components attached in parallel. Here $N$ is a random variable. If $X_1, X_2, \ldots$ denote the lifetime of the components of system 1 which are reported in a discrete scale, and in a same manner, $Y_1, Y_2, \ldots$ denote the lifetime of the components of system 2, then the lifetime of the two systems becomes $(X,Y)$.

\noindent {\sc Random Stress Model:} Suppose a system has two components. Each component is subject to random number of individual independent discrete stresses, say $\{X_1,X_2,...\}$ and $\{Y_1, Y_2, \ldots \}$, respectively. If $N$ is the number of stresses, then the observed stresses at the two components are
$X=\max\{X_1,...,X_N\}$ and $Y=\max\{Y_1,...,Y_N\}$, respectively.

The joint CDF of a bivariate GDGE distribution for $x \ge 0$ and $y \ge 0$ can be obtained as
\beanno
F_{X,Y}(x,y)&=&\sum_{n=1}^\infty P(X\leq x, Y\leq y|N=n)P(N=n) \\
&=& \theta \sum_{n=1}^\infty (1-p_1^{[ x ]+1})^{n\alpha_1}(1-p_2^{[ y ]+1})^{n\alpha_2}(1-\theta)^{n-1} \\
& = & \frac{\theta(1-p_1^{[ x ]+1})^{\alpha_1}(1-p_2^{[ y ]+1})^{\alpha_2}}{1-(1-\theta)(1-p_1^{[ x ]+1})^{\alpha_1}(1-p_2^{[ y ]+1})^{\alpha_2}}.
\eeanno
The above bivariate distribution with the parameter vector $\boldsymbol{\Omega}=(\alpha_1,\alpha_2,p_1,p_2,\theta)^T$ will be denoted
by BGDGE$(\alpha_1,\alpha_2,p_1,p_2,\theta)$.

It is interesting to note that if $\theta=1$, then we have
$$F_{X,Y}(x,y)=F_{DGE}(x;\alpha_1,p_1)F_{DGE}(y;\alpha_2,p_2),$$
i.e., $X$ and $Y$ become independent. Therefore, the parameter $\theta$ plays the role of the correlation parameter.
The joint PMF of $(X,Y)$ for $x\in \mathbb{N}_0$ and $y\in \mathbb{N}_0$ can be obtained as
$$f_{X,Y}(x,y)=F_{X,Y}(x,y)-F_{X,Y}(x-1,y)-F_{X,Y}(x,y-1)+F_{X,Y}(x-1,y-1).$$
More precisely, the joint PMF is given by
\begin{eqnarray}
f_{X,Y}(x,y)=g(x,y;\boldsymbol{\Omega})-g(x,y-1;\boldsymbol{\Omega}),    \label{bv-jpmf}
\end{eqnarray}
where
{\footnotesize\begin{eqnarray}\nonumber
g(x,y;\boldsymbol{\Omega})=\frac{\theta F_{DGE}(y;\alpha_2,p_2)f_{DGE}(x;\alpha_1,p_1)}{\left[1-(1-\theta)F_{DGE}(x;\alpha_1,p_1)F_{DGE}(y;\alpha_2,p_2)\right]
\left[1-(1-\theta)F_{DGE}(x-1;\alpha_1,p_1)F_{DGE}(y;\alpha_2,p_2)\right]}.
\end{eqnarray}}

Let us recall the bivariate geometric generalized exponential (BGGE) distribution as introduced by Kundu (2015).  Suppose $U_1, U_2,
\ldots$ are i.i.d. GE$(\alpha_1, \lambda_1)$ random variables, $V_1, V_2, \ldots$ are i.i.d. GE$(\alpha_2, \lambda_2)$ random variables,
$N$ is GM$(\theta)$ random variable and all are independently distributed.  The bivariate absolute continuous random variable $(U, V)$,
where
$$
U = \max\{U_1, \ldots, U_N\} \ \ \ \hbox{and} \ \ \ \ V = \max\{V_1, \ldots, V_N\}
$$
is said to have BGGE distribution with parameters $\alpha_1, \alpha_2, \lambda_1, \lambda_2$ and  $\theta$.  From now on it will be denoted by
BGGE$(\alpha_1, \alpha_2, \lambda_1, \lambda_2, \theta)$.  It can be shown similarly as Theorem 2 that if $(U, V) \sim$
BGGE$(\alpha_1, \alpha_2, \lambda_1, \lambda_2, \theta)$, then $(X, Y) \sim$ BGDGE$(\alpha_1,\alpha_2,p_1,p_2,\theta)$,
where $X = [U], Y = [V]$, $p_1 = e^{-\lambda_1}$ and $p_2 = e^{-\lambda_2}$.  Therefore, it is clear that the bivariate GDGE distribution is
the natural discrete version of the BGGE distribution.

Generation from a bivariate GDGE distribution is also quite straight forward.  First $N$ is generated from a GM$(\theta)$,
and once $N=n$ is observed, $U$ and $V$, can be generated from GE$(n\alpha_1,\lambda_1)$ and GE$(n\alpha_2,\lambda_2)$, respectively.
Then $([U], [V])$ becomes the required bivariate random variable.  Now, we have the following results for the bivariate GDGE distribution.

The PGF for $|u| < 1, |v| < 1$, and MGF for $t_1 < -\ln p_1, t_2 < -\ln p_2$ of $(X,Y)$ can be obtained as
$$
G_{X,Y}(u,v) = \theta \sum_{k=0}^{\infty} \sum_{j=1}^{\infty} \sum_{i=1}^{\infty} (-1)^{i+j}
{\alpha_1(k+1)\choose{j}}{\alpha_2(k+1)\choose{i}} (1-\theta)^k \frac{1-p_1^j}{1-up_1^j}\frac{1-p_2^i}{1-vp_2^i}.
$$
and
\beanno
M_{X,Y}(t_1,t_2)&=&E(e^{t_1X+t_2Y})=E_N\left\{E_{X,Y|N}(e^{t_1X+t_2Y})|N=n\right\}\\
&=&\theta\sum_{n=1}^\infty \sum_{j=1}^\infty\sum_{k=1}^\infty(-1)^{j+k}{n\alpha_1 \choose j}{n\alpha_2 \choose k}\frac{(1-p_1^j)(1-p_2^k)}{(1-p_1^je^{t_1})(1-p_2^ke^{t_2})},
\eeanno
respectively.

\noindent {\sc Theorem 4:} If $(X,Y)\sim$ BGDGE$(\alpha_1,\alpha_2,p_1,p_2,\theta)$, then we have the following results:

(a) $X\sim$ UGDGE$(\alpha_1,p_1,\theta)$ and  $Y\sim$ UGDGE$(\alpha_2,p_2,\theta)$.

(b) $X\leq x|Y\leq y\sim$ UGDGE$(\alpha_1,p_1,p^*)$, where $p^*=1-(1-\theta)(1-p_2^{y+1})^{\alpha_2}$.

(c) If $p_1=p_2=p$, then $\max\{X,Y\}\sim$ UGDGE$(\alpha_1+\alpha_2,p,\theta)$.

(d) $P(X\leq x|Y=y)=\frac{\left[1-(1-\theta)F_{DGE}(y;\alpha_2,p_2)\right]\left[1-(1-\theta)F_{DGE}(y-1;\alpha_2,p_2)\right][F_{X,Y}(x,y)-F_{X,Y}(x,y-1)]}{\theta f_{DGE}(y;\alpha_2,p_2)}$

\noindent {\sc Proof.} The proof of (a)-(c) is straight forward and we provide the proof of (d) as follows:
\begin{eqnarray}\nonumber
P(X\leq x|Y=y)&=&\sum_{n=1}^\infty P(X\leq x|Y=y, N=n)P(N=n|Y=y)\\\nonumber
&=&\sum_{n=1}^\infty (1-p_1^{x+1})^{n\alpha_1}P(N=n|Y=y).
\end{eqnarray}
By substitution the $P(N=n|Y=y)$, given by (19), the result is obtained. \qed

The joint PMF of $(X,Y,N)$ is also given by
\begin{eqnarray}\nonumber
f_{X,Y,N}(x,y,n)=\theta(1-\theta)^{n-1}\left[(1-p_1^{x+1})^{n\alpha_1}-(1-p_1^{x})^{n\alpha_1}\right]
\left[(1-p_2^{y+1})^{n\alpha_2}-(1-p_2^{y})^{n\alpha_2}\right].
\end{eqnarray}
Therefore, we see that
\bea
P(N=n|X=x,Y=y)&=&\left[(1-p_1^{x+1})^{n\alpha_1}-(1-p_1^{x})^{n\alpha_1}\right]
\left[(1-p_2^{y+1})^{n\alpha_2}-(1-p_2^{y})^{n\alpha_2}\right]  \nonumber \\
&\times& \frac{\theta(1-\theta)^{n-1}}{g(x,y;\boldsymbol{\Omega})-g(x,y-1;\boldsymbol{\Omega})},  \label{probn}
\eea
and hence we obtain the following conditional expectation,
\beanno
E(N|X=x,Y=y)&=&\frac{F_{X,Y}(x,y)a(x,y)-F_{X,Y}(x-1,y)a
(x-1,y)}{g(x,y;\boldsymbol{\Omega})-g(x,y-1;\boldsymbol{\Omega})}\\
&-&\frac{F_{X,Y}(x,y-1)a(x,y-1)-F_{X,Y}(x-1,y-1)a(x-1,y-1)}{g(x,y;\boldsymbol{\Omega})-g(x,y-1;\boldsymbol{\Omega})},
\eeanno
where $a(x,y)=\left[1-(1-\theta)(1-p_1^{x+1})^{\alpha_1}(1-p_2^{y+1})^{\alpha_2}\right]^{-1}$.

The BGDGE distribution, similar to its marginals, is closed under geometric maximum. More precisely, we have the following result whose proof is avoided.

\noindent {\sc Theorem 5:} Let $\{(U_i,V_i): i\geq1)\}$ be a sequence of i.i.d. BGDGE$(\alpha_1, \alpha_2, p_1, p_2, \theta)$ random variables, and $M\sim$ GM$(q)$, $0<q<1$. In addition, $M$ is independent of $(U_i,V_i)$'s. If we consider the random variables
\begin{eqnarray}
U=\max\{U_1,U_2,...,U_M\} \quad \rm and \quad V=\max\{V_1,V_2,...,V_M\},
\end{eqnarray}
then $(U,V)\sim$ BGDGE$(\alpha_1, \alpha_2, p_1, p_2, q\theta)$.

%\subsection{\sc BGDGE: Dependence Properties}

\section{\sc Statistical Inference}

\subsection{\sc Maximum Likelihood Estimation}

In this section we consider the maximum likelihood estimation of the unknown parameters of a BGDGE$(\alpha_1, \alpha_2, p_1, p_2, \theta)$
model based on a random sample of size $m$, namely ${\cal D} = \{(x_1, y_1), \ldots, (x_m,y_m)\}$.  The proposed bivariate GDGE model
has five parameters.  It is observed that the MLEs of the unknown parameters can be obtained by solving
a five-dimensional optimization problem.  To avoid that we have used an EM algorithm and in this case it is observed that at each `E'-step
the corresponding `M'-step can be obtained by maximizing two two-dimensional optimization problems.  Hence, the implementation of the
proposed EM algorithm is quite simple in practice.  The same EM algorithm with obvious modifications can also be used to compute the MLEs of
the unknown parameters of a UGDGE($\alpha, \theta, p)$ model.  We will indicate that towards the end of this section.

From now on we will indicate the parameter vector by ${\ve \Omega} = (\alpha_1, \alpha_2, p_1, p_2, \theta)$.
Based on the random sample ${\cal D}$ as mentioned above and using (\ref{bv-jpmf}), the log-likelihood function can be written as
\be
l({\ve \Omega}|{\cal D}) = \sum_{i=1}^m \ln \left \{g(x_i,y_i; {\ve \Omega}) - g(x_i, y_i-1; {\ve \Omega})\right \}.   \label{lg-lk}
\ee
Here $g(x,y)$ is same as defined in (\ref{bv-jpmf}).  The MLEs of the unknown parameters can be obtained by maximizing (\ref{lg-lk})
with respect to ${\ve \Omega}$.  It needs solving five non-linear equations
\be
\dot{l}_{\alpha_1}({\ve \Omega}|{\cal D}) = 0, \ \ \ \dot{l}_{\alpha_2}({\ve \Omega}|{\cal D}) = 0, \ \ \ \dot{l}_{p_1}({\ve \Omega}|{\cal D})
= 0, \ \ \ \dot{l}_{p_2}({\ve \Omega}|{\cal D}) = 0, \ \ \ \dot{l}_{\theta}({\ve \Omega}|{\cal D}) = 0,
\ee
simultaneously.  Clearly, they cannot be obtained in explicit forms.  Newton-Raphson method may be used to solve these non-linear
equations but it has the usual problem of convergence and the choice of the initial guesses.

To avoid that problems we propose to use an EM algorithm to compute the MLEs of the unknown parameters.  The main idea about the EM
algorithm is based on the following observations.  It is observed that if along with $(x,y)$, the associated $n$ is also known then the
MLEs of ${\ve \Omega}$ can be obtained in a more convenient manner.  Suppose we have the following observations
${\cal D}^* = \{(x_1, y_1,n_1), \ldots, (x_m, y_m, n_m)\}$.  Then based on ${\cal D}^*$, the log-likelihood function without the additive
constant can be written as
\be
l_{complete}({\ve \Omega}|{\cal D}^*) = m ln \theta + (k-m) \ln (1-\theta) + g_1(\alpha_1, p_1) + g_2(\alpha_2, p_2).  \label{com-ll}
\ee
Here $\ds k = \sum_{i=1}^m n_i$ and
$$
g_1(\alpha_1, p_1) = \sum_{i=1}^m \ln \left [ (1-p_1^{x_i+1})^{n_i \alpha_1} - (1-p_1^{x_i})^{n_i \alpha_1} \right ],
$$
$$
g_2(\alpha_2, p_2) = \sum_{i=1}^m \ln \left [ (1-p_2^{y_i+1})^{n_i \alpha_2} - (1-p_2^{y_i})^{n_i \alpha_2} \right ].
$$
The following result will be useful for further development.

\noindent {\sc Lemma 1:} (a) For any $0 < p_1 < 1$, $g_1(\alpha_1, p_1)$ is a unimodal function of $\alpha_1$.
(b) For any $0 < p_2 < 1$, $g_2(\alpha_2, p_2)$ is a unimodal function of $\alpha_2$.

\noindent {\sc Proof:} See in the Appendix.

Therefore, based on the ${\cal D}^*$, the MLE of $\theta$ can be obtained as
\be
\widehat{\theta} = \frac{m}{k}, \label{mle-th}
\ee
and the MLEs of $\alpha_1$ and  $p_1$ can be obtained by maximizing $g_1(\alpha_1, p_1)$ with respect to
the unknown parameters.  Similarly,
the MLEs of $\alpha_2$ and $p_2$ also can be obtained from $g_2(\alpha_2, p_2)$.  The maximization of $g_1(\alpha_1, p_1)$ and $g_2(\alpha_2,
p_2)$ can be obtained by using the profile likelihood method.  Therefore, based on the complete data ${\cal D}^*$ the MLEs can be
obtained quite conveniently and we will denote them as $\widehat{\ve \Omega} = (\widehat{\alpha}_1, \widehat{\alpha}_2, \widehat{p}_1,
\widehat{p}_2, \widehat{\theta})$.

Now, to implement the EM algorithm we treat this problem as a missing value problem.  It is assumed that the complete data set
is ${\cal D}^*$ and the observed data set is ${\cal D}$.  Therefore, for each $(x_i, y_i)$ the associated $n_i$ is missing in this case, and we need to
estimate $n_i$ from the observed data.  Let us use the following notations.  At the $j$th iterate of the EM algorithm the estimates of the parameters are
denoted by ${\ve \Omega}^{(j)} = (\alpha_1^{(j)}, \alpha_2^{(j)}, p_1^{(j)}, p_2^{(j)}, \theta^{(j)})$. At the $j$th iteration step the missing
$n_i$ is estimated by maximizing the conditional probability $P(N=n|X = x_i, Y=y_i)$ as in (\ref{probn}).  To compute $P(N=n|X = x_i, Y=y_i)$,
the parameter vector on the right hand side is replaced by ${\ve \Omega}^{(j)}$.  We denote the estimated $n_i$ as $\widetilde{n}_i^{(j)}$.
Therefore, we see that
\be
\widetilde{n}_i^{(j)} = \hbox{arg max}_n P(N=n|X = x_i, Y = y_i, {\ve \Omega}^{(j)}).    \label{exp-n}
\ee
Let us denote $\ds k^{(j)} = \sum_{i=1}^m \widetilde{n}_i^{(j)}$.  Now at the $(j+1)$th step the EM algorithm takes the following form.

\noindent `E'-Step: The `E'-step of the EM algorithm can be obtained by replacing $n_i$ with $\widetilde{n}_i^{(j)}$ in (\ref{com-ll}).
At this stage the {\it pseudo} log-likelihood function becomes
\be
l_{pseudo}({\ve \Omega}|{\cal D}^*,{\ve \Omega}^{(j)}) = m \ln \theta + (k^{(j)}-m) \ln (1-\theta) + g_1(\alpha_1, p_1|{\ve \Omega}^{(j)})
+ g_2(\alpha_2, p_2|{\ve \Omega}^{(j)}),    \label{pseudo-ll}
\ee
where
\be
g_1(\alpha_1, p_1|{\ve \Omega}^{(j)}) = \sum_{i=1}^m \ln \left [ (1-p_1^{x_i+1})^{\widetilde{n}_i^{(j)} \alpha_1} - (1-p_1^{x_i})^{\widetilde{n}_i^{(j)} \alpha_1} \right ],     \label{g1-func}
\ee
\be
g_2(\alpha_2, p_2|{\ve \Omega}^{(j)}) = \sum_{i=1}^m \ln \left [ (1-p_2^{y_i+1})^{\widetilde{n}_i^{(j)} \alpha_2} - (1-p_2^{y_i})^{\widetilde{n}_i^{(j)} \alpha_2} \right ],
\ee
and $\ds \widetilde{n}_i^{(j)}$ is obtained as in (\ref{exp-n}).

\noindent `M'-Step: The `M'-step involves maximizing (\ref{pseudo-ll}) with respect to the unknown parameters and they can be obtained as
follows: $\ds \theta^{(j+1)}  =  m/k^{j}$ and
\bea
(\alpha_1^{(j+1)}, p_1^{(j+1)}) & = & \hbox{arg max}_{(\alpha_1, p_1)} g_1(\alpha_1, p_1|{\ve \Omega}^{(j)}), \label{est-al1-p1} \\
(\alpha_2^{(j+1)}, p_2^{(j+1)}) & = & \hbox{arg max}_{(\alpha_2, p_2)} g_2(\alpha_2, p_2|{\ve \Omega}^{(j)}) \label{est-al2-p1}.
\eea

Now it is clear that the same EM algorithm can be used even for univariate GDGE distribution with obvious modification.  In this case
if we denote ${\ve \Omega} = (\alpha, p, \theta)$, ${\cal D}^* = \{(x_1, n_1), \ldots, (x_m, n_m)\}$,
$$
\widetilde{n}_i^{(j)} = \hbox{arg max}_n P(N=n|X = x_i, {\ve \Omega}^{(j)}),
$$
and all the other notations are same as before, then the EM algorithm takes the following form.

\noindent `E'-Step: The {\it pseudo} log-likelihood function becomes
\be
l_{pseudo}({\ve \Omega}|{\cal D}^*,{\ve \Omega}^{(j)}) = m \ln \theta + (k^{(j)}-m) \ln (1-\theta) + g_1(\alpha_1, p_1|{\ve \Omega}^{(j)}),
\label{ps-ll-1}
\ee
where $g_1(\alpha_1, p_1|{\ve \Omega}^{(j)})$ is same as defined in (\ref{g1-func}).

\noindent `M'-Step: The `M'-step involves maximizing (\ref{ps-ll-1}) with respect to the unknown parameters and they can be obtained as
follows: $\ds \theta^{(j+1)}  =  m/k^{j}$ and
$$
(\alpha_1^{(j+1)}, p_1^{(j+1)})  =  \hbox{arg max}_{(\alpha_1, p_1)} g_1(\alpha_1, p_1|{\ve \Omega}^{(j)}).
$$

\subsection{\sc Testing of Hypotheses}

In this section we discuss two different testing of hypotheses problems which have some practical relevance.  
In both the cases we have mainly used the likelihood ratio test (LRT).  In each case the MLE of any arbitrary parameter $\delta$
will be denoted by $\widehat{\delta}$ and under the null hypothesis it will be denoted by $\widetilde{\delta}$.

\noindent {\sc Test 1:} We want to test the following null hypothesis
\be
H_0: \alpha_1 = \alpha_2 = \alpha \hbox{ (unknown)} \ \ \hbox{and} \ \ p_1 = p_2 = p \hbox{ (unknown)}.
\ee
This is an important problem as it tests that the marginals are equal.  Under $H_0$, the MLEs of $\alpha, \theta$ and $p$ can be obtained
by using EM algorithm similarly as defined for bivariate GDGE distribution.  In this case at the `M'-step we need to maximize the
function
$$
g(\alpha, p|{\ve \Omega}^{(j)}) = \sum_{i=1}^m \ln \left [ (1-p^{x_i+1})^{\widetilde{n}_i^{(j)} \alpha} - (1-p^{x_i})^{\widetilde{n}_i^{(j)} \alpha} \right ],
$$
with respect to $\alpha$ and $p$.  Under $H_0$,
\be
2 (l(\widehat{\alpha}_1, \widehat{\alpha}_2, \widehat{p}_1, \widehat{p}_2, \widehat{\theta}| {\cal D})-
l(\widetilde{\alpha}, \widetilde{\alpha}, \widetilde{p}, \widetilde{p}, \widehat{\theta}| {\cal D})) \longrightarrow \chi^2_2,
\ee
see for example Casella and Berger (2001).

\noindent {\sc Test 2:} We want to test the following null hypothesis
\be
H_0: \theta = 1.
\ee
This is an important problem as it tests that the two marginals are independent and both of them follow univariate DGE distributions.
In this case the MLEs of $\alpha_1, \alpha_2, p_1$ and $p_2$ can also be obtained using the proposed EM algorithm with the obvious modification,
i.e. replacing $\widehat{\theta}$ = 1 at each stage.  Since $\theta$ is in the boundary, the standard asymptotic result does not work.
In this case, using Theorem 3 of Self and Liang (1987) yields that
\be
2 (l(\widehat{\alpha}_1, \widehat{\alpha}_2, \widehat{p}_1, \widehat{p}_2, \widehat{\theta}| {\cal D})-
l(\widetilde{\alpha}_1, \widetilde{\alpha}_2, \widetilde{p}_1, \widetilde{p}_2, 1| {\cal D})) \longrightarrow \frac{1}{2} + \frac{1}{2}
\chi^2_1.
\ee

\section{\sc Simulation and Data Analysis}

\subsection{\sc Simulation}

We have performed some simulation experiments to see how the proposed EM algorithm performs in computing the MLEs.
We have taken different sample sizes and two different $\theta$ values.  We have taken $\alpha_1 = \alpha_2 
= 2.0$, $p_1 = p_2 = 0.25$, $\theta$ = 0.25, 0.50, $n$ = 25, 50, 75 and 100.
In each case we have generated 
a random sample from the bivariate GDGE distribution with the given sample size and the parameter values.  We have estimated the 
parameters 
using the proposed EM algorithm.  We have reported the average estimates and the mean squared errors (MSEs) over 1000 
replications.  In each box the top figure represents the average estimate (AE) and the associated MSEs is reported below within a 
bracket.  The results are reported in Tables \ref{table-sim-1} to \ref{table-sim-2}.

\begin{table}[h]
\bc

\begin{tabular}{|l|c|c|l||c|c|}  \cline{1-6}
\hline
$n$ & $\alpha_1 = 2.0$ & $p_1 = 0.25$ & $\alpha_2 = 2.0$ & $p_2 = 0.25$ & $\theta = 0.25$   \\
   &   &   &   &   &   \\   \hline \hline
25 & 1.7124 & 0.1987 & 1.7215 & 0.1921 & 0.2011  \\
   & (0.5716) & (0.0581) & (0.5618) & (0.0534) & (0.0611) \\ \hline 
50 & 1.7618 & 0.2041 & 1.7691 & 0.2116 & 0.2278 \\
   & (0.3011) & (0.0312) & (0.2987) & (0.0349) & (0.0289) \\ \hline 
75 & 1.8312 & 0.2287 & 1.8579 & 0.2318 & 0.2410  \\
   & (0.2018) & (0.0211) & (0.2111) & (0.0228) & (0.0198) \\ \hline
100 & 1.9891 & 0.2510 & 2.0104 & 0.2498 & 0.2501 \\
    &(0.1439) & (0.0143) & (0.1411) & (0.0137) &(0.0114) \\ \hline
\end{tabular}
\ec
\caption{The AEs and the associated MSEs of the MLEs when $\alpha_1 = \alpha_2$ = 2.0, $p_1 = p_2$ = 0.25, $\theta$ = 0.25.
\label{table-sim-1}}
\end{table}

\begin{table}[h]
\bc
\begin{tabular}{|l|c|c|l||c|c|}  \cline{1-6}

\hline

$n$ & $\alpha_1 = 2.0$ & $p_1 = 0.25$ & $\alpha_2 = 2.0$ & $p_2 = 0.25$ & $\theta = 0.50$   \\
   &   &   &   &   &   \\   \hline \hline
25 & 1.7422 & 0.2012 & 1.7519 & 0.2065 & 0.2117  \\
   & (0.5218) & (0.0487) & (0.5198) & (0.0446) & (0.0576) \\ \hline 
50 & 1.7776 & 0.2145 & 1.7890 & 0.2208 & 0.2365 \\
   & (0.2567) & (0.0265) & (0.2514) & (0.0276) & (0.0245) \\ \hline 
75 & 1.9676 & 0.2376 & 1.9786 & 0.2406 & 0.2473  \\
   & (0.1676) & (0.0167) & (0.1632) & (0.0187) & (0.0141) \\ \hline
100 & 2.0015 & 0.2504 & 2.0011 & 0.2501 & 0.2500 \\
    &(0.1256) & (0.0110) & (0.1198) & (0.0116) &(0.0101) \\ \hline
\end{tabular}
\ec
\caption{The AEs and the associated MSEs of the MLEs when $\alpha_1 = \alpha_2$ = 2.0, $p_1 = p_2$ = 0.25, $\theta$ = 0.50.
\label{table-sim-2}}
\end{table}

Some of the points are quite clear from the simulation experiments that as the sample size increases the biases and MSEs decrease
in each case.  Moreover, it is also observed that as $\theta$ increases the biases and MSEs decrease for each estimators.

\subsection{\sc Data Analysis}

In this section we would present the analysis of a bivariate data set to show how the proposed model and the EM algorithm
work in practice.  This bivariate data set represents Italian Series A football match score data between Italian giants
`ACF Firontina' ($X_1$) and `Juventus' ($X_2$) during the time period 1996 to 2011.  The data set is available in Lee and Cha (2015)
and it is presented in Table \ref{football-data} for easy reference.  It is presented in the contingency table form
in Table \ref{data-cont}.

\begin{table}[h]
\bc
\begin{tabular}{|l|c|c|l||c|c|}  \cline{1-6}

\hline
Obs. & ACF  & Juventus & Obs.   & ACF & Juventus  \\
     & Firontina  &    &    &    Firontina    &           \\
     &   ($X_1$)  & ($X_2$) &  &  ($X_1$)  & ($X_2$)    \\
   &   &   &   &   &   \\   \hline \hline
1 & 1 & 2 & 14 & 1 & 2    \\
2 & 0 & 0 & 15 & 1 & 1    \\
3 & 1 & 1 & 16 & 1 & 3    \\
4 & 2 & 2 & 17 & 3 & 3    \\
5 & 1 & 1 & 18 & 0 & 1    \\
6 & 0 & 1 & 19 & 1 & 1    \\
7 & 1 & 1 & 20 & 1 & 2    \\
8 & 3 & 2 & 21 & 1 & 0    \\
9 & 1 & 1 & 22 & 3 & 0    \\
10 & 2 & 1 & 23 & 1 & 2   \\
11 & 1 & 2 & 24 & 1 & 1   \\
12 & 3 & 3 & 25 & 0 & 1   \\
13 & 0 & 1 & 26 & 0 & 1   \\    \hline
\end{tabular}
\ec
\caption{Italian Series A data \label{football-data}}
\end{table}
\begin{table}[h]
\bc
\begin{tabular}{|c|c|l|c|c|c|}  \cline{1-6}

\hline
$X_1 \downarrow$ $X_2 \rightarrow$ & 0 & 1 & 2 & 3 & Total  \\  \hline
 0 & 1 & 5 & 0 & 0 & 6  \\  \hline
 1 & 1 & 7 & 5 & 1 & 14 \\  \hline
 2 & 0 & 1 & 1 & 0 & 2  \\  \hline
 3 & 1 & 0 & 1 & 2 & 4  \\  \hline
 Total & 3 & 13 & 7 & 3 & 26  \\  \hline
\end{tabular}
\ec
\caption{Italian Series A data \label{data-cont}}
\end{table}

First we would like to fit univariate GDGE distribution to both the marginals.  We have used the EM algorithm to compute the
MLEs.  We present the MLEs and the associated 95\% confidence intervals in Table \ref{table-umle}.  Now to see whether
univariate GDGE fits the marginals or not we have calculated the $\chi^2$ values and also the associated $p$-values for both
the marginals.  The results are presented in the same Table \ref{table-umle}.  Since the $p$-values are greater than 0.1 for
both the marginals we conclude that univariate GDGE fits both the marginals well.
\begin{table}[h]
\bc
\begin{tabular}{|c|c|c|c|c|c|c|}  \cline{1-7}

\hline
Variable & $\alpha$ & $\theta$ & $p$ & LL & $\chi^2$ & $p$-value  \\  \hline
$X_1$ & 4.6587 & 0.9987 & 0.2618 & -33.4193 & 6.1486 & $>$ 0.1 \\
  & $\mp$ 0.8756 & $\mp$ 0.0014 & $\mp$ 0.0541 & & &  \\  \hline
$X_2$ & 6.8029 & 0.3288 & 0.1683 & -31.8832 & 0.6853 & $>$ 0.1 \\
  & $\mp$ 1.1562 & $\mp$ 0.0087 & $\mp$ 0.0465 & & &  \\  \hline
\end{tabular}
\ec
\caption{MLEs of the unknown parameters and the associated 95\% confidence intervals \label{table-umle}}
\end{table}

Now we would like to fit the bivariate GDGE distribution to the bivariate data set.  We have used the EM algorithm for the bivariate
GDGE distribution proposed in the previous section.  Based on the fitted univariate GDGE marginals we have used the following
initial values for the unknown parameters:
$$
\alpha_1^{(0)} = 4.6587, \ \ \ p_1^{(0)} = 0.2618, \ \ \ \alpha_2^{(0)} = 6.8029, \ \ \ p_2^{(0)} = 0.1683, \ \ \theta^{(0)} = (0.9987+0.3288)/2.0 = 0.6638.
$$
The EM algorithm stops after nineteen iteration and the MLEs and the associated 95\% confidence bounds are presented within parenthesis
below
$$
\widehat{\alpha}_1 = 4.5519 (\mp 1.1101) \ \ \widehat{p}_1 = 0.2570 (\mp 0.0721)
$$
$$
\widehat{\alpha}_2 = 8.3892 (\mp 1.9767) \ \
\widehat{p}_2 = 0.2250 (\mp 0.0518)
$$
$$
\widehat{\theta} = 0.9211 (\mp 0.0312).
$$

Now to check whether the proposed BDGE fits the bivariate data or not we have obtained the observed and expected values and they are
presented in Table \ref{chisq}.  The chi-square value is 7.79 and the corresponding $p$-value with 9 degrees of freedom is greater than 0.1.  Hence
we cannot reject the hypothesis that the data are coming from a bivariate GDGE distribution.

\begin{table}[h]
\bc
\begin{tabular}{|c|c|l|c|c|}  \cline{1-5}

\hline
$X_1 \downarrow$ $X_2 \rightarrow$ & 0 & 1 & 2 & 3   \\  \hline
 0 & 1(0.64) & 5(3.31) & 0(1.75) & 0 (0.30)  \\  \hline
 1 & 1(1.17) & 7(6.32) & 5(3.51) & 1 (0.98)  \\  \hline
 2 & 0(0.88) & 1(2.67) & 1(1.54) & 0(0.44)   \\  \hline
 3 & 1(0.84) & 0(0.69) & 1(0.78) & 2(1.16)  \\  \hline
\end{tabular}
\ec
\caption{Observed and expected frequencies for Italian Series A data \label{chisq}}
\end{table}

\section{\sc Conclusions}

In this paper we have introduced univariate and bivariate GDGE distributions using the method proposed by Marshall and Olkin (1997). Apparently this is the first time the method of Marshall and Olkin (1997) has been used for the discrete bivariate distributions.
We have derived different properties of the proposed distributions.  It is observed that both the univariate and bivariate distributions are very flexible.  The MLEs of the unknown parameters cannot be obtained in explicit forms.  We have proposed to use a new EM algorithm which is applicable for the discrete distributions.  We have performed some simulation experiments to see the 
effectiveness of the proposed EM algorithm, and 
analyzed one data set for illustrative purposes.  It is observed that the proposed method and the new EM
algorithm work quite well in practice.

In this paper, we have considered the univariate and bivariate cases. It is important to see how it can be generalized to the multivariate case. More work is needed along this direction.

\section*{\sc Acknowledgements:} The authors would like to thank the reviewers and the associate editor for their constructive 
comments which have helped us to improve the manuscript significantly.

\section*{\sc Appendix}

\noindent {\sc PROOF OF LEMMA 1:}

It is enough to show that the function
$$
h(\alpha|p) = \ln \left [ (1-p^{j+1})^{\alpha} - (1-p^j)^{\alpha} \right ],
$$
for $0 < p < 1$, and for any $j = 0, 1, \ldots$, is a log-concave function of $\alpha$.  By straight forward calculation it can be seen that
$$
\frac{d^2}{d\alpha^2} h(\alpha|p) = \frac{h_1(\alpha|p)}{h_2(\alpha|p)}<0,
$$
where
\beanno
h_1(\alpha|p) & = & -(1-p^j)^{\alpha}(1-p^{j+1})^{\alpha} \left \{\ln(1-p^{j+1})-\ln(1-p^j)\right \}^2,   \\
h_2(\alpha|p) & = & \left \{(1-p^{j+1})^{\alpha}- (1-p^j)^{\alpha}\right \}^2.
\eeanno \qed

%\newpage

\end{document}